# Controlling Cherenkov angles with resonance transition radiation

Xiao Lin[1,2], Sajan Easo[3*], Yichen Shen[4], Hongsheng Chen[1], Baile Zhang[2,5], John D. Joannopoulos[4], Marin Soljačić[4], and Ido Kaminer[4,6]

[1]*State Key Laboratory of Modern Optical Instrumentation, The Electromagnetics Academy at Zhejiang University, Zhejiang University, Hangzhou 310027, China.*
[2]*Division of Physics and Applied Physics, School of Physical and Mathematical Sciences, Nanyang Technological University, Singapore 637371, Singapore.*
[3]*Particle Physics Department, Rutherford-Appleton Laboratory (STFC-UKRI), Didcot, OX110QX, United Kingdom.*
[4]*Department of Physics, Massachusetts Institute of Technology, Cambridge, MA 02139, USA.*
[5]*Centre for Disruptive Photonic Technologies, NTU, Singapore 637371, Singapore.*
[6]*Department of Electrical Engineering, Technion-Israel Institute of Technology, Haifa 32000, Israel*
[*]*email: Sajan.Easo@cern.ch*

Cherenkov radiation provides a valuable way to identify high energy particles in a wide momentum range, through the relation between the particle velocity and the Cherenkov angle. However, since the Cherenkov angle depends only on material's permittivity, the material unavoidably sets a fundamental limit to the momentum coverage and sensitivity of Cherenkov detectors. For example, Ring Imaging Cherenkov detectors must employ materials transparent to the frequency of interest as well as possessing permittivities close to unity to identify particles in the multi GeV range, and thus are often limited to large gas chambers. It would be extremely important albeit challenging to lift this fundamental limit and control Cherenkov angles as preferred. Here we propose a new mechanism that uses the constructive interference of resonance transition radiation from photonic crystals to generate both forward and backward Cherenkov radiation. This mechanism can control Cherenkov angles in a flexible way with high sensitivity to any desired range of velocities. Photonic crystals thus overcome the severe material limit for Cherenkov detectors, enabling the use of transparent materials with arbitrary values of permittivity, and provide a promising option suited for identification of particles at high energy with enhanced sensitivity.

The relation between the angle of Cherenkov radiation cones (denoted as Cherenkov angle $\theta$ below) [1,2] and the velocity $v$ of charged particles is of fundamental importance to many applications [3-6]. For example, this determines the sensitivity of different types of Cherenkov detectors such as the Ring Image Cherenkov (RICH) detectors [7, 8] for particle identification.

However, the relation between the Cherenkov angle and the particle velocity is inherently limited by the material in which the Cherenkov radiation is emitted. This unavoidably sets a strict limit on the design of Cherenkov detectors. For conventional Cherenkov radiation generated in a nonmagnetic material, when the particle velocity is known, the Cherenkov angle relies *only* on the material's relative permittivity $\varepsilon_r$ (which determines the refractive index $n = \sqrt{\varepsilon_r}$) through the formula $\cos\theta = (n\beta)^{-1}$, where $\beta = \frac{v}{c}$ and $c$ is the speed of light in free space. Regular transparent dielectrics are not suitable for conventional Cherenkov detectors. This is because these materials have a relative permittivity far above unity and the Cherenkov angle would saturate to a value independent of the particle velocity. For example, quartz has relative permittivity around 2 and therefore can be used only in the limited momentum range below 3.5 GeV/c (and even that requires using water instead of free space to out-couple the light) [9]. In order to distinguish between relativistic particles, Cherenkov detectors require the radiator materials to have a relative permittivity very close to unity. For example, gas radiators are typically used to detect particles with momentum higher than 10 GeV/c [10, 11], and aerogels with relative permittivity around 1.06 have been used for particle identification in parts of the 1-10 GeV/c momentum range [11, 12].

The limitation of Cherenkov radiation in regular transparent dielectrics also comes from another reason: the total internal reflection at the air-dielectric interface will prevent the Cherenkov radiation generated by relativistic particles from being observed for relative permittivity $\varepsilon_r > 2$, as coincidentally happens for most transparent dielectrics, and especially in the visible spectrum [13]. Having $\varepsilon_r > 2$ leads to the following inequality: $\lim_{v \to c} \left( \frac{\varepsilon_r \omega^2}{c^2} - \frac{\omega^2}{v^2} \right)^{1/2} > \frac{\omega}{c}$ for Cherenkov radiation fields inside the dielectric, and then the fields outside the dielectric become evanescent. Besides, material losses (bulk absorption) have a large impact on the performance of Cherenkov detectors. Recently, anisotropic metal-based metamaterials, which still requires one component of the relative permittivity very close to one, have been proposed to control Cherenkov angles [14]. However, the existence of a small loss, which is particularly unavoidable in metal-based system, will destroy the relation between the Cherenkov angle and the particle velocity [15]. Therefore, it shall be necessary to use purely transparent systems to gain efficient control of the Cherenkov angle.

The above facts severely limit the potential choice of materials for the design of Cherenkov detectors. Therefore, despite the long history of studies of Cherenkov radiation and its applications [3-6, 16-20], with much recent renewed interest and progress in the topic [21-30], the ability to control the Cherenkov angle in a flexible way is still limited by the permittivity of the radiator material and remains very challenging.

In this work, we propose a new underlying mechanism for the generation of Cherenkov radiation from a one-dimensional (1D) photonic crystal composed of widely-available transparent dielectrics, which can transmit into air and thus can be used in existing Cherenkov detector designs like the RICH. This comes from the constructive interference of the forward or backward resonance transition radiation from periodic dielectric interfaces. Therefore, along with the tremendous choice of lossless dielectrics and the flexibility in the design of periodic structures [31], this mechanism allows photonic crystals to flexibly control both the forward and backward Cherenkov angles. We note that while many phenomena of Cherenkov radiation have been studied in photonic crystals [23,28,32], including the backward (or reversed) Cherenkov radiation [21], spectroscopy of photonic nanostructures [3], and novel compact radiation sources [4,18-20], the possibility of using photonic crystals to tailor the Cherenkov angle has not been directly addressed. In addition, the proposed Cherenkov detectors based on photonic crystals are different from transition radiation detectors [33], where the latter relies only on the intensity of resonance transition radiation from multilayer structures and neglects the information of radiation angles [34-36].

To highlight the underlying physics, we begin by schematically showing in Fig. 1 that the effective Cherenkov radiation can be generated by the constructive interference of resonance transition radiation excited from multiple interfaces in 1D photonic crystals. For 1D photonic crystals, the simplest structure can be constructed by two different materials, such as two transparent dielectrics with relative permittivity denoted as $\varepsilon_{r1}$ and $\varepsilon_{r2}$. When the relativistic particle with a charge of $q$ and a velocity of $\bar{v} = \hat{z}\beta c$ penetrates through a 1D photonic crystal, the forward (backward) radiation can be generated in the bottom (top) air region, as shown in Fig. 1(a). As a conceptual demonstration of resonance transition radiation, figure 1(b) presents the radiation field distribution (without the charge field) generated from a swift electron ($\beta = 0.5022$) passing through a 1D photonic crystal ($\varepsilon_{r1} = 2$ and $\varepsilon_{r2} = 2.3$). Resonance transition radiation from 1D photonic crystals is analytically calculated by extending Ginzburg and Frank's theory of transition radiation [35,36] to a 1D photonic crystal structure; see detailed calculations in supplementary notes 1-2. Since the particle velocity is below the Cherenkov threshold, i.e. $\beta < [max(\varepsilon_{r1}, \varepsilon_{r2})]^{-1/2}$, there will be no conventional Cherenkov radiation within each dielectric. However, plane-like waves are still emitted within the photonic crystal near the particle trajectory, as Fig. 1(b) clearly shows. Interestingly, the $z$-component of the Poynting's vector $S$ (which represents the direction of power flow) is antiparallel to the direction of motion of the particle. Consequently, more radiation energy enters into the top air region than into the bottom air region. These are characteristic features of the effective backward Cherenkov radiation, which originates only from the constructive interference of resonance transition radiation in the backward direction. This new mechanism for the generation of Cherenkov radiation is different from that of conventional Cherenkov radiation described by the theory developed by Frank and Tamm [2, 34] and that of Smith-Purcell radiation [37, 38]. For the latter two cases, the generated fields are directly emitted into the air region without the intermediate modulation by a periodic dielectric environment, and the charged particle moves only within one material without crossing interfaces between different materials. Moreover, the Cherenkov angle from photonic crystals can be designed to be sensitive to relativistic velocities; see Figs. 2-4. For example, as shown in Figs. 1(c-g), when the value of $\beta$ of a charged particle increases from 0.99, 0.992, 0.994, 0.996, to 0.998, the corresponding backward Cherenkov angle in the top air region decreases from 10.8°, 9.7°, 8.4°, 6.9° to 4.8°. This demonstrates high angular sensitivity to small changes in the particle velocity, which is desired for particle identification from Cherenkov detectors.

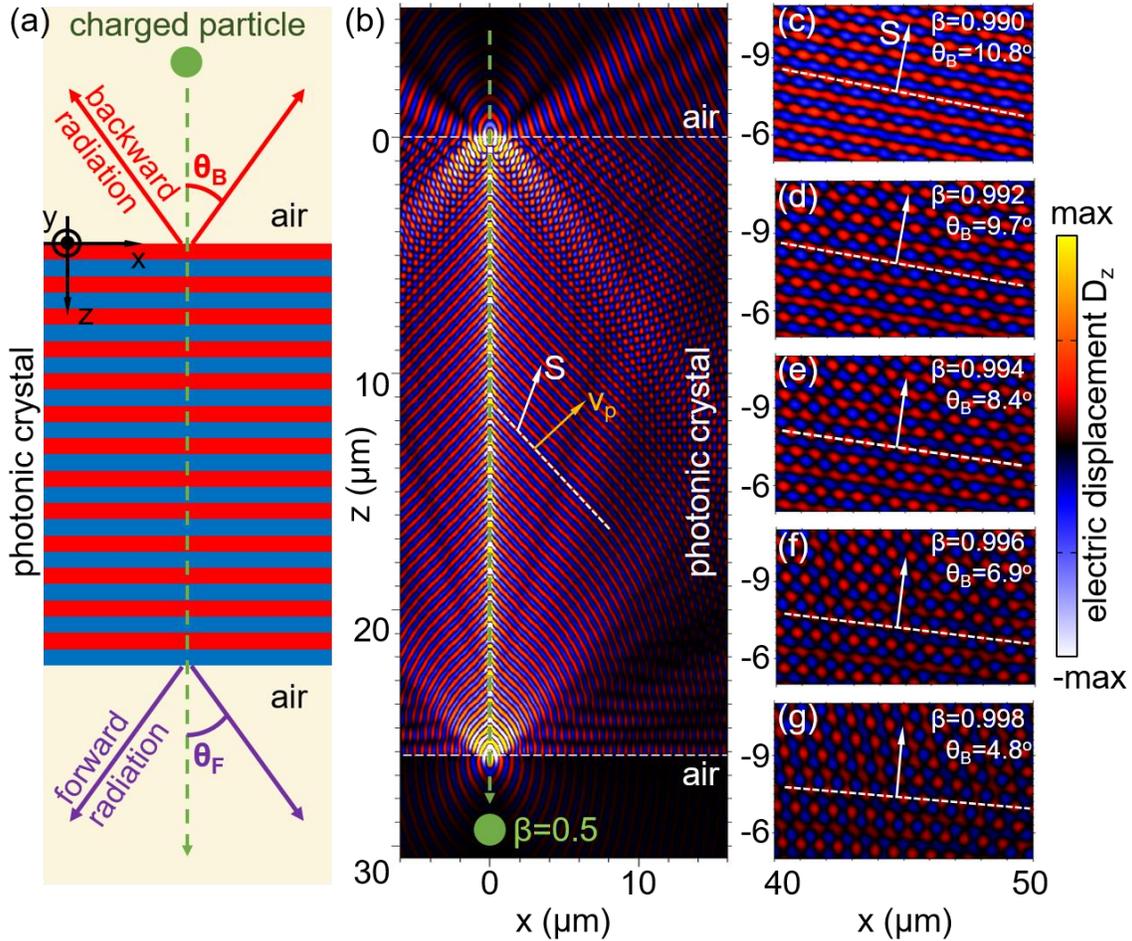

**Figure 1 | Schematic of controlling Cherenkov angles with photonic crystals.** (a) Structural schematic. The forward (backward) radiation is collected in the bottom (top) air region. (b-g) Field distribution of backward Cherenkov radiation induced by the constructive interference of resonance transition radiation in the backward direction. In (b), plane-like waves are excited near the particle trajectory (dashed green arrow), with the z-components of the Poynting's vector $S$ and phase velocity $v_p$ being both antiparallel to the direction of motion of the particle. In (c-g), the Cherenkov angle is shown by the phase fronts of the far field in the top air region, exhibiting high sensitivity to the particle velocity $v = \beta c$. Here, and in the figures below, the working wavelength in air $\lambda = 2\pi c/\omega$ is set to be 700 nm. In (b), the photonic crystal consists

of 40 unit cells; the thickness of the unit cell is $d_{unit} = 0.9346\lambda$; within each unit cell, the thicknesses for the two dielectric constituents are $d_1 = d_2 = 0.5d_{unit}$; $\varepsilon_{r1} = 2$ and $\varepsilon_{r2} = 2.3$. In (c-g), the thickness of the photonic crystal is 2 mm, with $d_{unit} = 0.2792\lambda$, $d_1 = 0.6d_{unit}$, $d_2 = 0.4d_{unit}$, $\varepsilon_{r1} = 10.6$ and $\varepsilon_{r2} = 2.1$.

The effective Cherenkov radiation induced by resonance transition radiation from 1D photonic crystals can be a new scheme to control Cherenkov angles. To gain an intuitive understanding of this scheme, we can qualitatively analyze the interaction between the charged particle and the eigenmode of photonic crystals. When the particle is assumed to move along the *z*-direction, it induces a current density of $\bar{J}^q(\bar{r},t) = \hat{z}\frac{qv}{2\pi\rho}\delta(z-vt)\delta(\rho)$ in cylindrical coordinates [35, 36, 39]. By transforming all quantities to the frequency domain, we have the particle-induced fields proportional to $\exp(i\frac{\omega}{v}z)$ [35, 36, 39], at each angular frequency $\omega$. From the momentum-matching condition, the charged particle is prone to excite the eigenmodes of photonic crystals with the wavevector along the *z*-direction being $k_z = \frac{c}{v} \cdot \frac{\omega}{c}$ [18]. Moreover, in order to guarantee that the excited modes inside the photonic crystal can reach the detectors, which are generally located at the air region, we need to avoid the total internal reflection at the photonic crystal-air interface. This requires the wavevector along the $\rho$-direction of the excited modes being $k_\rho \leq \frac{\omega}{c}$. Due to the momentum matching along the $\rho$-direction at the photonic crystal-air interface, $k_\rho$ of the excited modes propagating in the air region is the same as that in the photonic crystal. This way, $k_\rho$ determines the Cherenkov angle $\theta$ in the air region, since $k_\rho = \frac{\omega}{c}\sin(\theta)$.

There are two different schemes to control Cherenkov angles with photonic crystals, as shown in Fig. 2. From the above analysis, the relation between the Cherenkov angle in air and the particle velocity is effectively determined by the isofrequency contour of photonic crystals at each frequency, i.e., the relation between the wavevectors $k_z$ and $k_\rho$. Therefore, it is very straightforward to control the Cherenkov angle through tailoring the isofrequency contour. Two representative kinds of conceptual isofrequency contours of photonic crystals are shown in Fig. 2. When the range of the particle velocity is $[v_{min}, v_{max}]$, the corresponding range of the Cherenkov angle in air is $[\theta_{min}, \theta_{max}]$ in Fig. 2(a) but $[\theta_{max}, \theta_{min}]$ in Fig. 2(b), respectively. These two isofrequency contours in Fig. 2 can be treated as two common schemes to control the Cherenkov angle. While the first scheme in Fig. 2(a) is similar to that proposed in Ref. [14] with the use of anisotropic metamaterials, the second scheme in Fig. 2(b) has not been discussed in the context of angle control.

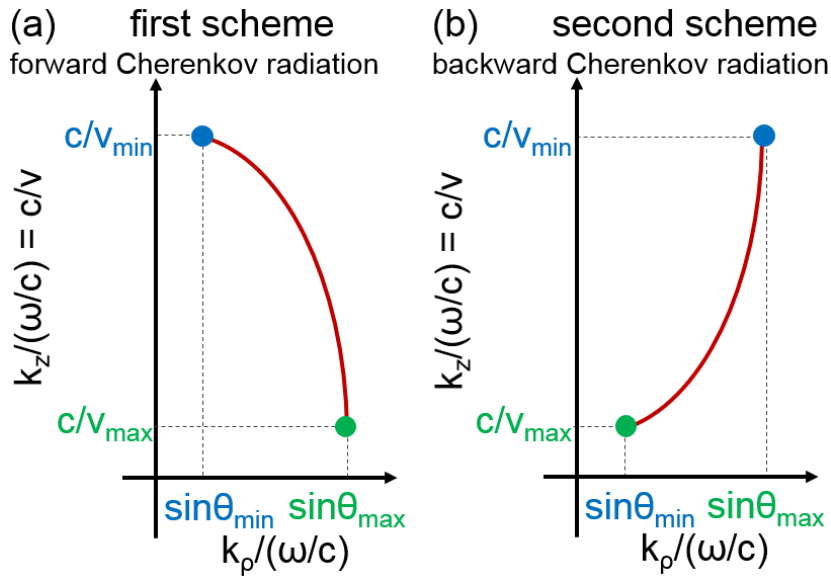

**Figure 2 | Two conceptual schemes of controlling forward and backward Cherenkov angles with photonic crystals.** Hypothetical isofrequency contours of photonic crystals, i.e. the relation between wavevectors $k_\rho$ and $k_z$, determines the relation between the Cherenkov angle in air ($\theta = asin(\frac{k_\rho}{\omega/c})$) and the particle velocity ($\frac{v}{c} = \frac{\omega/c}{k_z}$). The goal of the two schemes is to create a wide range of Cherenkov angles for a narrow range of the particle velocity. The maximum Cherenkov angle corresponds to the maximum particle velocity in the first scheme in (a), and corresponds to the minimum particle velocity in the second scheme in (b).

The two proposed schemes of controlling Cherenkov angles with photonic crystals in Fig. 2 are exemplified in Fig. 3 and Fig. 4. When a charged particle moves inside a bulk transparent dielectric, conventional Cherenkov radiation emerges with $k_\rho = \frac{\omega}{c}\sqrt{\varepsilon_r - \beta^{-2}}$ [34]. When $\varepsilon_r > 2$ and $\beta \to 1$, we have $k_\rho > \frac{\omega}{c}$. This indicates that the generated Cherenkov radiation inside dielectrics will be totally reflected at the dielectric-air interface. In the following, the working wavelength in air is set to be 700 nm, and the relative permittivity of the two dielectric constituents for the designed photonic crystals is set to be $\varepsilon_{r1} = 10.6$ (such as GaP) and $\varepsilon_{r2} = 2.1$ (SiO$_2$) [13], respectively. This way, conventional Cherenkov radiation generated inside each dielectric in photonic crystals will not contribute to the radiation in the air region but will be guided within the waveguide-like photonic crystals.

Figure 3 exemplifies the first scheme of controlling Cherenkov angles proposed in Fig. 2(a). In order to demonstrate the relation between Cherenkov angles and the particle velocity achieved from the photonic crystal, figures 3(a,b) show the angular spectral energy density from the photonic crystal in the forward and backward directions, respectively. The forward (backward) angular spectral energy density $U(\lambda, \theta) = \frac{dW}{d\omega d\Omega}$ [35, 36, 39], which represents the energy $W$ radiated per unit angular frequency $\omega$ per unit solid angle $\Omega$ ($d\Omega = 2\pi sin\theta d\theta$), characterizes the forward (backward) radiation in the bottom (top) air region; see detailed calculation in supplementary note 3. All the radiation energy in the air region flows predominantly along paths such that in each case the corresponding radiation angle is highly dependent on the particle velocity; see Figs. 3(a,b). Moreover, the forward radiation energy in Fig. 3(a) is ~100 times larger than the backward radiation energy in Fig. 3(b). Therefore, the resonance transition radiation from this particular example of the photonic crystal can be effectively considered as the forward Cherenkov radiation. The weak backward radiation in Fig. 3(b) is attributed to the destructive interference of resonance transition radiation in the backward direction.

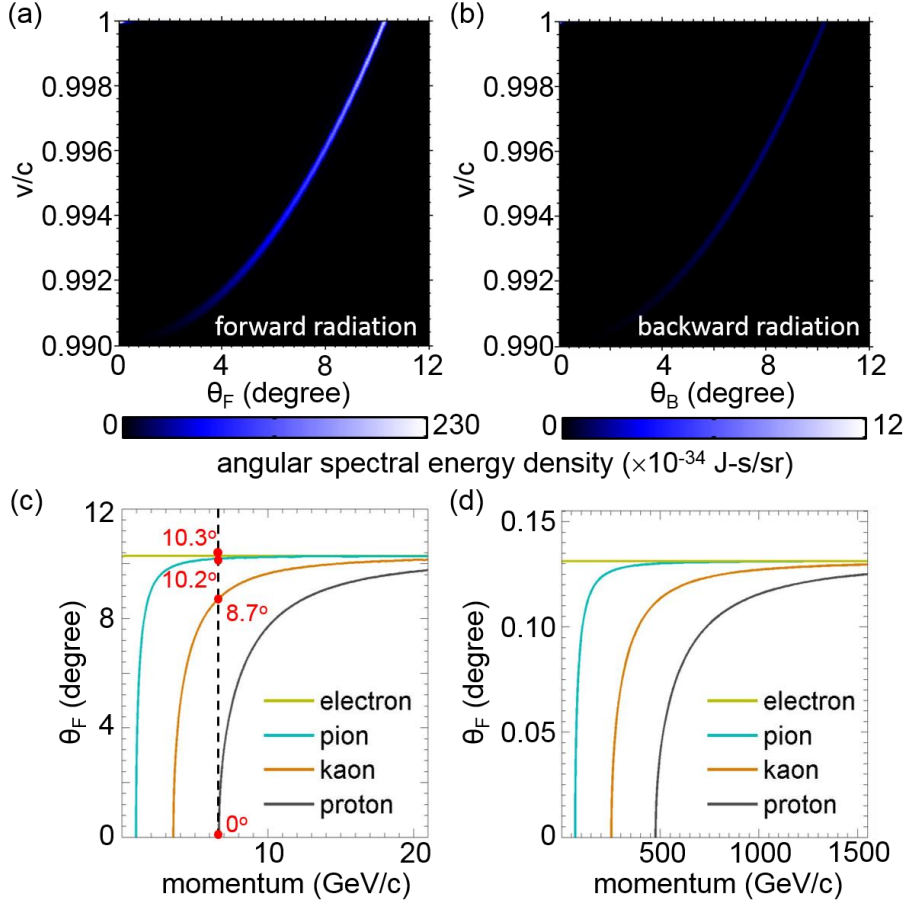

**Figure 3 | Controlling Cherenkov angles with photonic crystals by using our first proposed scheme.**
(a-b) Angular spectral energy density of forward (backward) radiation in the bottom (top) air region. The highly directional radiation in (a) shows the relation between the Cherenkov angle and the particle velocity. (c) Cherenkov angles versus the particle momenta for four kinds of particles, where the velocity in (a) is translated to the momentum using the masses of different charged particles. The thickness of the photonic crystal is 2 mm, with $d_{unit} = 1.0205\lambda$, $d_1 = 0.3d_{unit}$, $d_2 = 0.7d_{unit}$, $\varepsilon_{r1} = 10.6$ and $\varepsilon_{r2} = 2.1$. (d) Cherenkov angles versus the particle momenta, where $d_{unit} = 1.0144\lambda$ and the other parameters are the same as those in (c). The results in this figure follow the proposed scheme in Fig. 2(a).

The Cherenkov angle from photonic crystals is well-suited for high-energy particle identification. For example, by applying the angle-velocity relation of Fig. 3(a), we show the relation between the particle
9

momentum and the Cherenkov angle for four kinds of particles with different masses in Fig. 3(c). As an example, for particles with a momentum of 6.6 GeV/c, the Cherenkov angles corresponding to electron, pion, kaon and proton hypothesis are 10.3º, 10.2º, 8.7º and 0º respectively. This would enable the different particle types to be effectively distinguished from one another in the region near this momentum. Conventional radiators cannot cover this range because the permittivity of gas is too low and the permittivities of solid transparent materials (e.g. quartz) are too high. There is a dearth of suitable materials with the required permittivity (around ~1.06): silica aerogels have been typically used for this momentum range, but they suffer from significant losses due to Rayleigh scattering. In addition, the Cherenkov angle from photonic crystals can be tailored to be well-suited for particle identification for other momentum ranges. For example, Fig. 3(d) shows the use of photonic crystals for particle identification with extremely high momenta (~500 GeV/c), which cannot be achieved with conventional radiator materials, as it requires for example a gas at extremely low density which is difficult to use in a Cherenkov detector.

While the forward Cherenkov radiation has been extensively studied [9, 10], the backward Cherenkov radiation has never been considered for the design of Cherenkov detectors. For the identification of charged particles, the backward Cherenkov radiation has a distinct advantage over the forward Cherenkov radiation [40]: since the emitted photon and the particle are naturally separated in opposite directions, their physical interference is minimized. This also can lead to Cherenkov detector designs with two radiators for two different momentum ranges where the emitted photons go in forward or backward directions depending on the momentum of the charged particle. This would reduce the hit occupancy in the corresponding photon detector planes and thus improve the particle identification performance, compared to the configuration where all the photons go forward and reach a single photon detector plane. We find that photonic crystals can also be used to control the backward Cherenkov angles, as shown in Fig. 4.

Figure 4 shows the second scheme of controlling Cherenkov angles proposed in Fig. 2(b). From the angular spectral energy density, Figs. 4(a,b) show that all radiation energy in the air region predominantly goes in directions such that for each direction the corresponding radiation angle is sensitive to the particle velocity; the backward radiation energy in Fig. 4(b) is ~10 times larger than the forward radiation energy in Fig. 4(a). This is attributed to the constructive (destructive) interference of resonance transition radiation in the backward (forward) direction. Therefore, the resonance transition radiation in Fig. 4 is effectively treated as the backward Cherenkov radiation. As a schematic demonstration of backward Cherenkov angles, the far field distribution from a swift electron with different velocities highlighted by yellow dots in Fig. 4(b) are shown by the respective Figs. 1(c-g). Here the maximum particle velocity corresponds to the minimum Cherenkov angle, opposite to the previous scheme shown in Fig. 3 and to that of conventional radiator materials. Fig. 4(c) shows the relation between the Cherenkov angles and the particle momenta for a range of higher momenta. For example, for particles with a momentum of 20 GeV/c, the Cherenkov angles corresponding to electron, pion, kaon and proton hypothesis are 0º, 0.48º, 1.88º and 3.58º respectively. These different Cherenkov angles indicate that this second scheme is also suitable for identification of high energy particles, and there is no fundamental limit to the range of momenta that a photonic crystal can cover when using a design based on this scheme.

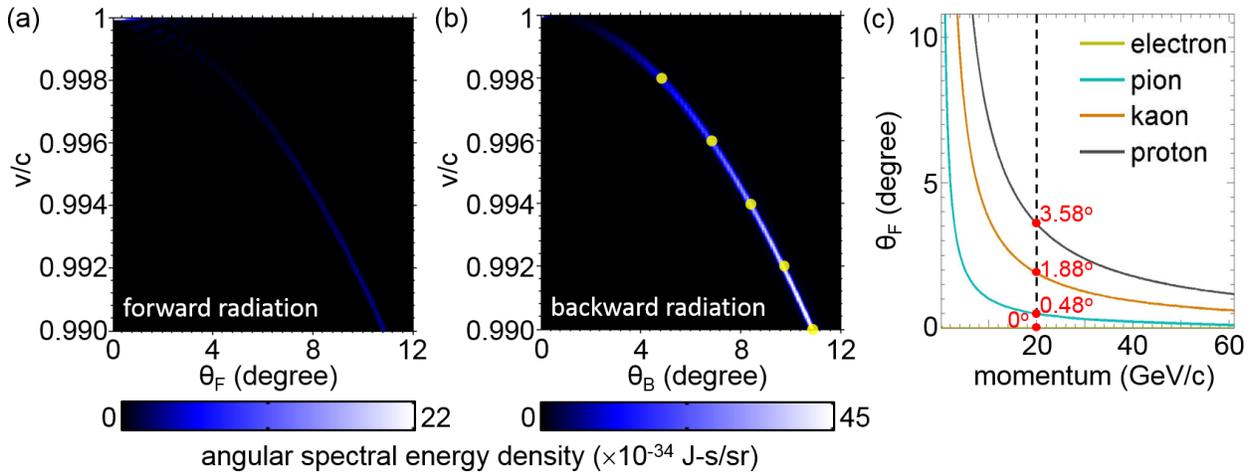

**Figure 4 | Controlling Cherenkov angles with photonic crystals by using our second proposed scheme.** (a-b) Angular spectral energy density of forward (backward) radiation in the bottom (top) air region. The highly directional radiation in (b) shows the relation between the Cherenkov angle and the particle velocity. Cherenkov angles at five different particle velocities, denoted as yellow dots in (b), are schematically shown by the far field radiation in the top air region in Figs. 1(c-g), respectively. (c) Cherenkov angles versus the

particle momenta for four kinds of particles, where the velocity in (b) is translated to the momentum using the different masses of the particles. The thickness of the photonic crystal is 2 mm, with $d_{unit} = 0.2792\lambda$, $d_1 = 0.6 d_{unit}$, $d_2 = 0.4 d_{unit}$, $\varepsilon_{r1} = 10.6$ and $\varepsilon_{r2} = 2.1$. The results in this figure follow the proposed scheme in Fig. 2(b).

From above analysis, it follows that photonic crystals can offer more freedom to control Cherenkov angles. This is because photonic crystals enable the controlling of both the forward and backward Cherenkov angles (see Figs. 3&4), while conventional radiator materials, such as aerogels, gases, or anisotropic metal-based metamaterials, can control only the forward Cherenkov angle [9,10,14]. Moreover, photonic crystals does not have strict requirement on the value of permittivities of dielectric constituents, which thus overcomes the material limit, enabling the use of transparent materials with arbitrary values of permittivity for the design of Cherenkov detectors.

In order to facilitate the potential design of Cherenkov detectors based on photonic crystals, some of the salient features of the effective Cherenkov radiation from photonic crystals are described below. The number of photons emitted per unit length from the photonic crystal described above is similar to that from ideally-lossless anisotropic metamaterials in Ref.[14], but is one order of magnitude smaller than what could be achieved from an isotropic material of hypothetical similar refractive index; see Figs. 3(a),S3(a)&S5(b). From a practical perspective, one may improve the total photon yield through further structural optimization, through increasing the thickness of photonic crystals, or through the using of 2D or 3D photonic crystals [31]. Using 2D and 3D photonic crystals also suggests the possibility of additional mechanisms for the enhancement of Cherenkov radiation, such as the excitation of modes with high density of states at points of van-Hove singularities [21, 41].

Notice that the prospect of a thin photonic structure for the design of Cherenkov detectors is highly desirable because practical Cherenkov-based detectors for particles above 10 GeV/c momentum typically require gas radiators, which are typically at least one meter long. Using photonic crystals the radiator length may be reduced to few millimeters, considering that such crystals will already have ~10000 periods, with the potential to create very efficient emission. The constraint in increasing the layers is that the fraction of radiation length introduced by the photonic crystal should be kept small enough so that the charged particles create only negligible amount of secondary particles when traversing the photonic crystal; there are many transparent dielectric materials made from combinations of low Z materials like silicon, oxygen, carbon, and nitrogen, which have a low rate of secondary emissions. In addition, we note that using photonic crystals to design Cherenkov detectors might suffer from the chromatic aberration induced by the periodic structure. This also happens for Cherenkov detectors from anisotropic metal-based metamaterials [14], due to the high frequency dispersion of the permittivity in metals. The chromatic aberration will cause the Cherenkov angle to be sensitive to the frequency. One solution is to limit the frequency range that is detected by the photon detectors, using filters. The resulting reduction of the number of photons detected may be compensated by increasing the thickness of the photonic crystal with the constraint to keep the low rate of secondary particle production. There are also other directions one can consider to broaden the frequency bandwidth. One is to design additional optical elements or layers directly on the photonic crystal. A more fundamental approach, which fixes the chromatic aberration directly inside the photonic crystal, is to use materials with anomalous dispersion [42] to construct photonic crystals, since the material's anomalous dispersion can help to cancel the dispersion problems caused by the periodic structure.

Before closing, it is important to note that typical Cherenkov detectors receive particles along different trajectories, and therefore one cannot assume that all of the particles move along the $z$-direction. Instead, in experiments like the Large Hadron Collider (LHC), several high energy particles are produced in the region near the nominal interaction point of the beam particles, and then they move outward from that region. For example, the particles produced in the pseudorapidity range from 2 to 5 (i.e, travelling within a cone of about 300 mrad around the axis of collision), would not be incident at normal angle on a flat plane of the photonic crystal. For the tracks incident at such small angles, corrections can be applied to the above formalism [35] to estimate the Cherenkov angle. One can also design the crystal such that it forms a spherical segment to facilitate normal incidence of a large fraction of such particles so that the formalism developed here can be used directly. Since the radius of curvature of such a segment is much larger than the structural periodicity, we can consider the curved crystal as effectively a planar structure to a very good approximation. The fabrication of such a structure is different from that of conventional 1D photonic crystals. Nevertheless, there already exists a number of viable approaches one can consider: multilayer polymer sheets are being used as 1D photonic crystals in a number of applications [43, 44] ; such structure can be easily bent. Moreover, there are a range of approaches for flexible photonic crystals [45-47]. Finally, conventional layer deposition methods and optical lithography methods could be modified and re-optimized to work on curved (e.g., spherical) surfaces.

To conclude, this work introduces a new mechanism, i.e. the constructive interference of resonance transition radiation in the forward or backward direction, to generate the Cherenkov radiation from a 1D photonic crystal. This new mechanism allows to control both the forward and backward Cherenkov angles in a flexible way, despite using transparent dielectrics with their relative permittivities far above unity, and thus overcomes the severe material limit for the design of conventional Cherenkov detectors. With the combined advantages of the abundant choice of dielectrics and the flexibility in the structural design, photonic crystals thus provide a new viable platform for the design of Cherenkov detectors with enhanced sensitivity and for the design of novel radiation sources.

**Acknowledgements.** This work was sponsored by the National Natural Science Foundation of China (Grants No. 61625502, 61574127 and 61601408), the ZJNSF (LY17F010008), the Top-Notch Young Talents Program of China, the Fundamental Research Funds for the Central Universities, the Innovation Joint Research Center for Cyber-Physical-Society System, Nanyang Technological University for NAP Start-Up Grant, the Singapore Ministry of Education (Grant No. MOE2015-T2-1-070, MOE2011-T3-1-005, and Tier 1 RG174/16 (S)), and the US Army Research Laboratory and the US Army Research Office through the Institute for Soldier Nanotechnologies (Contract No. W911NF-13-D-0001). M. Soljačić was supported in part (reading and analysis of the manuscript) by the MIT S3TEC Energy Research Frontier Center of the Department of Energy under Grant No. DESC0001299. I. Kaminer was partially supported by the Seventh Framework Programme of the European Research Council (FP7-Marie Curie IOF) under Grant No. 328853-MC-BSiCS. S.Easo is a staff member of STFC (which is a constituent part of UKRI) in U.K.

# Supplementary Information for

# "Controlling Cherenkov angles with resonance transition radiation"

**Supplementary note 1: Transition radiation from a single interface**

We begin with the analysis of transition radiation from a charged particle perpendicularly crossing a single interface between two different isotropic regions; see Supplementary Fig. 1. The swift particle with a charge of $q$ propagates along the $+z$ direction with a velocity of $v$. The two regions are denoted as region $j$ and region $j+1$, and have a relative permittivity of $\varepsilon_{r,j}$ and $\varepsilon_{r,j+1}$, respectively, where $j$ is a positive integer. The corresponding plane of the interface between these two regions is at $z = d_j$. The detailed analytical calculation of transition radiation for the case with $d_j = 0$ has been extensively studied [23,35,36,39], including that in our recent work [39]. For the convenience of following discussions, we briefly introduce the calculation procedure of transition radiation for the case with arbitrary $d_j$ below [39].

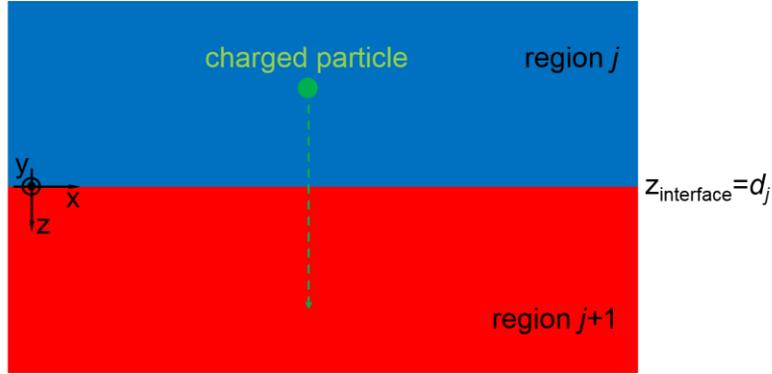

**Supplementary Figure 1 | Structural schematic of transition radiation when a charged particle perpendicularly crosses a single interfaces.**

Within the framework of classical electrodynamics, the current density induced by a swift charged particle is [35,36]:

$$\bar{J}^q(\bar{r},t) = \hat{z}qv\delta(x)\delta(y)\delta(z - vt) \tag{1}$$

By decomposing all the quantities in Fourier components in time and in the coordinates $\bar{r}_\perp = \hat{x}x + \hat{y}y$ perpendicular to the moving charge's trajectory, one has

$$\bar{J}^q(\bar{r},t) = \hat{z}J_z^q(\bar{r},t) = \hat{z}\int j_{\bar{\kappa}_\perp,\omega}^q(z)e^{i(\bar{\kappa}_\perp\cdot\bar{r}_\perp-\omega t)}d\bar{\kappa}_\perp d\omega \tag{2}$$

$$\bar{E}(\bar{r},t) = \int \bar{E}_{\bar{\kappa}_\perp,\omega}(z)e^{i(\bar{\kappa}_\perp\cdot\bar{r}_\perp-\omega t)}d\bar{\kappa}_\perp d\omega, \quad \bar{H}(\bar{r},t) = \int \bar{H}_{\bar{\kappa}_\perp,\omega}(z)e^{i(\bar{\kappa}_\perp\cdot\bar{r}_\perp-\omega t)}d\bar{\kappa}_\perp d\omega \tag{3}$$

where $\bar{\kappa}_\perp = \hat{x}\kappa_x + \hat{y}\kappa_y$. From equations (1-2), one obtains $j_{\bar{\kappa}_\perp,\omega}^q(z) = \frac{q}{(2\pi)^3}e^{i\frac{\omega}{v}z}$. Below we will mainly use the fields (such as $j_{\bar{\kappa}_\perp,\omega}^q$, $\bar{E}_{\bar{\kappa}_\perp,\omega}$ and $\bar{H}_{\bar{\kappa}_\perp,\omega}(z)$) in the Fourier decomposition. For the sake of simplicity, we will not give the indices $\bar{\kappa}_\perp$ and $\omega$ for the corresponding Fourier components. By solving Maxwell equations, one can find the following equation for $E_z$ (the component along the charge's trajectory):

$$\frac{\partial^2}{\partial z^2}(\varepsilon_r E_z) + \varepsilon_r\left(\frac{\omega^2}{c^2}\varepsilon_r - \kappa_\perp^2\right)E_z = -\frac{i\omega\mu_0 q}{(2\pi)^3}\left(\varepsilon_r - \frac{c^2}{v^2}\right)e^{i\frac{\omega}{v}z} \tag{4}$$

where $\varepsilon_r$ is the relative permittivity (i.e. $\varepsilon_r = \varepsilon_{r,j}$ or $\varepsilon_{r,j+1}$); $\varepsilon_0$ and $\mu_0$ is the permittivity and the permeability of free space, respectively; $c$ is the speed of light in free space. Equation (4) can be solved in each region, and the solutions should be joined by matching the boundary conditions at $z = d_j$, i.e.

$$\hat{n}\times(\bar{E}_{\perp,j} - \bar{E}_{\perp,j+1})|_{z=d_j} = 0, \quad \hat{n}\times(\bar{H}_{\perp,j} - \bar{H}_{\perp,j+1})|_{z=d_j} = 0 \tag{5}$$

where $\hat{n} = -\hat{z}$, and $\bar{J}_s$ is the surface current density. Such a solution will be a sum of a field induced by the charge in a homogeneous medium ($E_z^q$) and the freely radiated field ($E_z^R$) [35,36], i.e. $E_z = E_z^q + E_z^R$, where

$$E_{z,j}^q = \frac{-iq}{\omega\varepsilon_0(2\pi)^3}\frac{1-\frac{c^2}{v^2\varepsilon_{r,j}}}{(\varepsilon_{r,j}-\frac{c^2}{v^2}-\frac{\kappa_\perp^2 c^2}{\omega^2})}e^{i\frac{\omega}{v}z} \tag{6-1}$$

$$E_{z,j+1}^q = \frac{-iq}{\omega\varepsilon_0(2\pi)^3}\frac{1-\frac{c^2}{v^2\varepsilon_{r,j+1}}}{(\varepsilon_{r,j+1}-\frac{c^2}{v^2}-\frac{\kappa_\perp^2 c^2}{\omega^2})}e^{i\frac{\omega}{v}z} \tag{6-2}$$

$$E_{z,j}^R = \frac{iq}{\omega\varepsilon_0(2\pi)^3}\cdot a_{j|j+1}^-\cdot e^{-ik_{z,j}z} \tag{7-1}$$

$$E_{z,j+1}^R = \frac{iq}{\omega\varepsilon_0(2\pi)^3}\cdot a_{j|j+1}^+\cdot e^{+ik_{z,j+1}z} \tag{7-2}$$

where the component of the wavevectors for the radiated fields along the $z$ direction are $k_{z,j} = \sqrt{\frac{\varepsilon_{r,j}\omega^2}{c^2} - \kappa_\perp^2}$ and $k_{z,j+1} = \sqrt{\frac{\varepsilon_{r,j+1}\omega^2}{c^2} - \kappa_\perp^2}$, respectively. Both the backward radiated field in equation (7-1) and the forward radiated field in equation (7-2) propagate away from the interface. By matching the boundary conditions, one can obtain the two radiation factors $a^-_{j|j+1}$ and $a^+_{j|j+1}$ for the radiation fields in region $j$ and region $j+1$, i.e.

$$a^-_{j|j+1} = a^{-,o}_{j|j+1} e^{+ik_{z,j}d_j} e^{i\frac{\omega}{v}d_j} \tag{8}$$

$$a^+_{j|j+1} = a^{+,o}_{j|j+1} e^{-ik_{z,j+1}d_j} e^{i\frac{\omega}{v}d_j} \tag{9}$$

$$a^{-,o}_{j|j+1} = \frac{\frac{v}{c}\frac{\kappa_\perp^2 c^2}{\omega^2 \varepsilon_{r,j}}(\varepsilon_{r,j+1}-\varepsilon_{r,j})(1-\frac{v^2}{c^2}\varepsilon_{r,j}+\frac{v}{c}\frac{k_{z,j+1}}{\omega/c})}{(1-\frac{v^2}{c^2}\varepsilon_{r,j}+\frac{\kappa_\perp^2 v^2}{\omega^2})(1+\frac{v}{c}\frac{k_{z,j+1}}{\omega/c})[\varepsilon_{r,j}\frac{k_{z,j+1}}{\omega/c}+\varepsilon_{r,j+1}\frac{k_{z,j}}{\omega/c}]} \tag{10}$$

$$a^{+,o}_{j|j+1} = \frac{\frac{v}{c}\frac{\kappa_\perp^2 c^2}{\omega^2 \varepsilon_{r,j+1}}(\varepsilon_{r,j+1}-\varepsilon_{r,j})(1-\frac{v^2}{c^2}\varepsilon_{r,j+1}-\frac{v}{c}\frac{k_{z,j}}{\omega/c})}{(1-\frac{v^2}{c^2}\varepsilon_{r,j+1}+\frac{\kappa_\perp^2 v^2}{\omega^2})(1-\frac{v}{c}\frac{k_{z,j}}{\omega/c})[\varepsilon_{r,j}\frac{k_{z,j+1}}{\omega/c}+\varepsilon_{r,j+1}\frac{k_{z,j}}{\omega/c}]} \tag{11}$$

The factors of $a^{-,o}_{j|j+1}$ in equation (10) and $a^{+,o}_{j|j+1}$ in equation (11) correspond to $a^-_{j|j+1}$ in equation (8) and $a^+_{j|j+1}$ in equation (9) for the case with $d_j = 0$, respectivley.

The fields in equation (3), expressed in the Cartesian coordinates $(x, y, z)$, can also be transformed into the cylindrical coordinates $(\rho, \phi, z)$. After some calculations, one has

$$\bar{E}_j(\bar{r},t) = \bar{E}^q_j(\bar{r},t) + \bar{E}^R_j(\bar{r},t) \tag{12}$$

$$\bar{E}^q_j(\bar{r},t) = \hat{z}\int_{-\infty}^{+\infty} d\omega \frac{-q}{8\pi\omega\varepsilon_0\varepsilon_{r,j}}\left(\frac{\omega^2}{c^2}\varepsilon_{r,j} - \frac{\omega^2}{v^2}\right) H_0^{(1)}\left(\rho\sqrt{\frac{\omega^2}{c^2}\varepsilon_{r,j} - \frac{\omega^2}{v^2}}\right)e^{i(\frac{\omega}{v}z-\omega t)} +$$

$$\hat{\rho}\int_{-\infty}^{+\infty} d\omega \frac{-q}{8\pi\omega\varepsilon_0\varepsilon_{r,j}}(i\frac{\omega}{v})(-\sqrt{\frac{\omega^2}{c^2}\varepsilon_{r,j} - \frac{\omega^2}{v^2}}H_1^{(1)}(\rho\sqrt{\frac{\omega^2}{c^2}\varepsilon_{r,j} - \frac{\omega^2}{v^2}}))e^{i(\frac{\omega}{v}z-\omega t)} \tag{13}$$

$$\bar{E}^R_j(\bar{r},t) = \hat{z}\int_{-\infty}^{+\infty} d\omega \int_0^{+\infty} d\kappa_\perp \cdot \frac{iq}{\omega\varepsilon_0(2\pi)^3} a^-_{j|j+1}\kappa_\perp(2\pi J_0(\kappa_\perp\rho))e^{i[-k_{z,j}z-\omega t]} +$$

$$\hat{\rho}\int_{-\infty}^{+\infty} d\omega \int_0^{+\infty} d\kappa_\perp \cdot \frac{iq}{\omega\varepsilon_0(2\pi)^3} a^-_{j|j+1}k_{z,j}(i2\pi J_1(\kappa_\perp\rho))e^{i[-k_{z,j}z-\omega t]} \tag{14}$$

$$\bar{E}_{j+1}(\bar{r},t) = \bar{E}^q_{j+1}(\bar{r},t) + \bar{E}^R_{j+1}(\bar{r},t) \tag{15}$$

$$\bar{E}^q_{j+1}(\bar{r},t) = \hat{z}\int_{-\infty}^{+\infty}d\omega\frac{-q}{8\pi\omega\varepsilon_0\varepsilon_{r,j+1}}\left(\frac{\omega^2}{c^2}\varepsilon_{r,j+1}-\frac{\omega^2}{v^2}\right)H_0^{(1)}\left(\rho\sqrt{\frac{\omega^2}{c^2}\varepsilon_{r,j+1}-\frac{\omega^2}{v^2}}\right)e^{i\left(\frac{\omega}{v}z-\omega t\right)}+$$

$$\hat{\rho}\int_{-\infty}^{+\infty}d\omega\frac{-q}{8\pi\omega\varepsilon_0\varepsilon_{r,j+1}}(i\frac{\omega}{v})(-\sqrt{\frac{\omega^2}{c^2}\varepsilon_{r,j+1}-\frac{\omega^2}{v^2}}H_1^{(1)}(\rho\sqrt{\frac{\omega^2}{c^2}\varepsilon_{r,j+1}-\frac{\omega^2}{v^2}}))e^{i\left(\frac{\omega}{v}z-\omega t\right)} \tag{16}$$

$$\bar{E}^R_{j+1}(\bar{r},t) = \hat{z}\int_{-\infty}^{+\infty}d\omega\int_0^{+\infty}d\kappa_\perp\cdot\frac{iq}{\omega\varepsilon_0(2\pi)^3}a^+_{j|j+1}\kappa_\perp\left(2\pi J_0(\kappa_\perp\rho)\right)e^{i[+k_{z,j+1}z-\omega t]}+$$

$$\hat{\rho}\int_{-\infty}^{+\infty}d\omega\int_0^{+\infty}d\kappa_\perp\cdot\frac{iq}{\omega\varepsilon_0(2\pi)^3}a_2(-k_{z,j+1})(i2\pi J_1(\kappa_\perp\rho))e^{i[+k_{z,j+1}z-\omega t]} \tag{17}$$

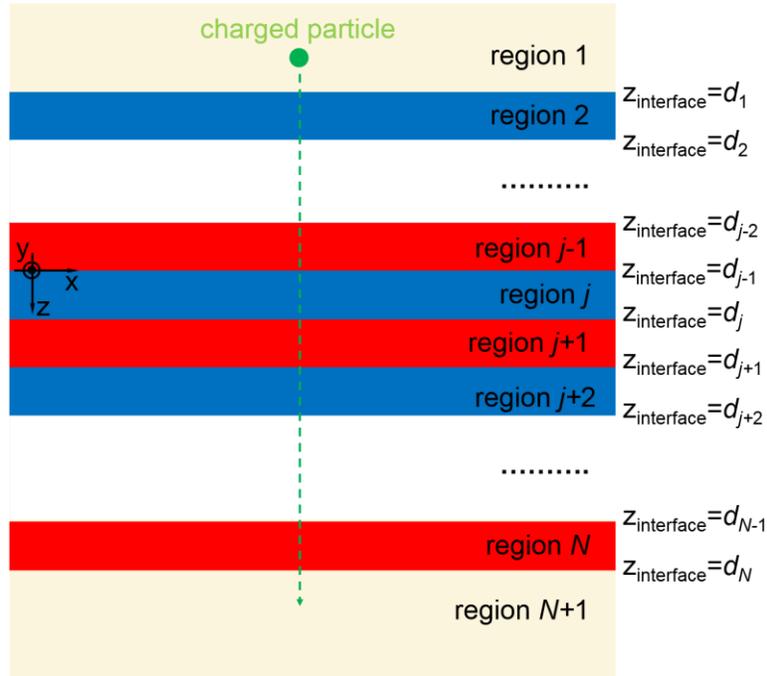

**Supplementary Figure 2 | Structural schematic of resonance transition radiation when a charged particle perpendicularly crosses multiple interfaces.**

### Supplementary note 2: Resonance transition radiation from multiple interfaces

When a charged particle crosses a periodic structure, the presence of periodicity leads, naturally, to the coherence of emitted waves that appears at different interfaces, and as a consequence, to resonance effects [35,36]. Therefore, transition radiation from a periodic structure is called the resonance transition radiation [38,39,48]. By extending Ginzburg and Frank's theory on transition radiation to one-dimensional photonic crystals, we can analytically solve the resonance transition radiation when a charged particle perpendicularly crosses multiple interfaces; see Supplementary Fig. 2.

As a general analysis, we consider the studied system having $N+1$ regions and $N$ interfaces; see Supplementary Fig. 2. Region 1 and region $N+1$ correspond to the top and bottom air regions, respectively. The other setup are the same as those in supplementary note 1. Since the emitted fields of transition radiation at each interface can transmit into each region, the emitted fields in each region shall be a summation of transition radiation from each interface. Actually, transition radiation at each interface can be equivalently treated as the secondary radiation source, where its corresponding field distribution in the whole system is required to be solved. Below we demonstrate the main calculation procedure by separately solving the field distribution in each region from the forward transition radiation at the $j-1|j$ interface (i.e. the interface between region $j-1$ and region $j$) and from the backward transition radiation at the $j|j+1$ interface. The main procedure is similar to the calculation of the reflection/transmission in multilayer systems [49].

*Part one: Field distribution in region j from the forward transition radiation at the $j-1|j$ interface*

For the forward transition radiation at the $j-1|j$ interface, from equation (9), we have the forward radiation factor as follows:

$$a_{j-1|j}^+ = a_{j-1|j}^{+,o} e^{-ik_{z,j}d_{j-1}} e^{i\frac{\omega}{v}d_{j-1}} \tag{18}$$

From equation (7-2), the radiated field in region $j$ becomes:

$$E_{z,j}^R = A_{j,o}^+ [e^{+ik_{z,j}(z-d_{j-1})} + \tilde{R}_{j|j+1} e^{-ik_{z,j}(z-d_{j-1})} \cdot e^{2ik_{z,j}(d_j-d_{j-1})}] \tag{19}$$

$$A_{j,o}^+ = \frac{iq}{\omega\varepsilon_0(2\pi)^3} \cdot a_{j-1|j}^{+,o} e^{i\frac{\omega}{v}d_{j-1}} \cdot M_j \tag{20}$$

$$M_j = \frac{1}{1-\tilde{R}_{j|j+1}\tilde{R}_{j|j-1}e^{2ik_{z,j}(d_j-d_{j-1})}} \tag{21}$$

In above equations, the factor $M_j$ characterize the multiple reflections at the $j-1|j$ and $j|j+1$ interfaces; $\tilde{R}_{j|j+1}$ and $\tilde{R}_{j|j-1}$ are the *generalized reflection coefficients* of transverse-magnetic (TM, or *p*-polarized) waves for the multi-layer system at the $j|j+1$ and $j|j-1$ interfaces, respectively; see more introduction about the generalized reflection coefficient in Ref.[49]. In the subscript of $\tilde{R}_{j|j+1}$ and $\tilde{R}_{j|j-1}$, the TM plane wave is incident from region $j$ (the first number) and transmitted to region $j+1$ or $j-1$ (the second number), respectively. This rule applies for other reflection and transmission coefficients of TM waves in the following. From Ref.[49], we have

$$\tilde{R}_{j|j+1} = R_{j|j+1} + \frac{T_{j|j+1}\tilde{R}_{j+1|j+2}T_{j+1|j}e^{2ik_{z,j+1}(d_{j+1}-d_j)}}{1-R_{j+1|j}\tilde{R}_{j+1|j+2}e^{2ik_{z,j+1}(d_{j+1}-d_j)}} \tag{22-1}$$

$$\tilde{R}_{j|j-1} = R_{j|j-1} + \frac{T_{j|j-1}\tilde{R}_{j-1|j-2}T_{j-1|j}e^{2ik_{z,j-1}(d_{j-1}-d_{j-2})}}{1-R_{j-1|j}\tilde{R}_{j-1|j-2}e^{2ik_{z,j-1}(d_{j-1}-d_{j-2})}} \tag{22-2}$$

where $\tilde{R}_{N|N+1} = R_{N|N+1}$ and $\tilde{R}_{2|1} = R_{2|1}$; $R_{j|j+1} = -R_{j+1|j} = \frac{\frac{k_{z,j}}{\varepsilon_{r,j}} - \frac{k_{z,j+1}}{\varepsilon_{r,j+1}}}{\frac{k_{z,j}}{\varepsilon_{r,j}} + \frac{k_{z,j+1}}{\varepsilon_{r,j+1}}}$ are the reflection coefficient; $T_{j|j+1} = \frac{2\frac{k_{z,j}}{\varepsilon_{r,j}}}{\frac{k_{z,j}}{\varepsilon_{r,j}} + \frac{k_{z,j+1}}{\varepsilon_{r,j+1}}} \cdot \frac{\varepsilon_{r,j}}{\varepsilon_{r,j+1}}$ and $T_{j+1|j} = \frac{2\frac{k_{z,j+1}}{\varepsilon_{r,j+1}}}{\frac{k_{z,j+1}}{\varepsilon_{r,j+1}} + \frac{k_{z,j}}{\varepsilon_{r,j}}} \cdot \frac{\varepsilon_{r,j+1}}{\varepsilon_{r,j}}$ are the transmission coefficients. It is worthy to note that in this work, the generalized reflection coefficients and other reflection/transmission coefficients are defined for the $E_z$ field (see equation (19)), instead of the magnetic field. For TM waves, while the reflection coefficients for the $E_z$ field and the magnetic field are the same (i.e. $R^{E_z}_{j|j+1} = R^{H}_{j|j+1}$), the transmission coefficients for the $E_z$ field and the magnetic field are different (i.e. $T^{E_z}_{j|j+1} = T^{H}_{j|j+1} \cdot \frac{\varepsilon_{r,j}}{\varepsilon_{r,j+1}}$).

*Part two: Field distribution in region $m$ ($m > j$) from the forward transition radiation at the $j - 1|j$ interface*

Since the forward transition radiation at the $j-1|j$ interface propagates along the $+z$ direction, part of the emitted fields will transmit into region $m$ ($m > j$). These transmitted fields in region $m$ can be obtained by following the thought of calculating the reflection/transmission in multilayer systems [49]. The field in region $m$ can be expressed as:

$$E^R_{z,m} = C^+_{j|m}[e^{+ik_{z,m}(z-d_{j-1})} + \tilde{R}_{m|m+1}e^{-ik_{z,m}(z-d_{j-1})} \cdot e^{2ik_{z,m}(d_m-d_{j-1})}] \tag{23}$$

In equation (23), only the factor $C_{j|m}^+$ is unknown. When $m = j + 1$ in equation (23), at the $j|j+1$ interface with $z = d_j$, we have

$$C_{j|j+1}^+ e^{+ik_{z,j+1}(d_j-d_{j-1})} = A_{j,o}^+ e^{+ik_{z,j}(d_j-d_{j-1})} \cdot S_{j|j+1} \tag{24}$$

$$S_{j|j+1} = \frac{T_{j|j+1}}{1-R_{j+1|j}\tilde{R}_{j+1|j+2}e^{2ik_{z,j+1}(d_{j+1}-d_j)}} \tag{25}$$

Namely, the value of $C_{j|j+1}^+$ is determined by $A_{jo}^+$ and the transmission at the $j|j+1$ interface. By following the operation in equation (24), we have

$$C_{j|j+2}^+ e^{+ik_{z,j+2}(d_{j+1}-d_{j-1})} = C_{j|j+1}^+ e^{+ik_{z,j+1}(d_{j+1}-d_{j-1})} \cdot S_{j+1|j+2} \tag{26}$$

$$C_{j|j+3}^+ e^{+ik_{z,j+3}(d_{j+2}-d_{j-1})} = C_{j|j+2}^+ e^{+ik_{z,j+2}(d_{j+2}-d_{j-1})} \cdot S_{j+2|j+3} \tag{27}$$

······

$$C_{j|m-1}^+ e^{+ik_{z,m-1}(d_{m-2}-d_{j-1})} = C_{j|m-2}^+ e^{+ik_{z,m-2}(d_{m-2}-d_{j-1})} \cdot S_{m-2|m-1} \tag{28}$$

$$C_{j|m}^+ e^{+ik_{z,m}(d_{m-1}-d_{j-1})} = C_{j|m-1}^+ e^{+ik_{z,m-1}(d_{m-1}-d_{j-1})} \cdot S_{m-1|m} \tag{29}$$

From equations (24-29), one has

$$C_{j|m}^+ e^{+ik_{z,m}(d_{m-1}-d_{j-1})} = \tilde{T}_{j|m}^{m>j} \cdot A_{j,o}^+ e^{+ik_{z,j}(d_{j-1}-d_{j-1})} \tag{30}$$

$$\tilde{T}_{j|m}^{m>j} = \prod_{n=j}^{n=m-1} S_{n|n+1} e^{+ik_{z,n}(d_n-d_{n-1})} \tag{31}$$

$\tilde{T}_{j|m}^{m>j}$ can be treated as the *generalized transmission coefficient* from region $j$ to region $m$ [49].

*Part three: Field distribution in region $m$ ($m < j$) from the forward transition radiation at the $j - 1|j$ interface*

Equation (19) can be equivalently transformed to following expression:

$$E_{z,j}^R = B_{j,o}^+ e^{+ik_{z,j}(z-d_{j-1})} + AR_{j,o}^-[e^{-ik_{z,j}(z-d_{j-1})} + \tilde{R}_{j|j-1}e^{+ik_{z,j}(z-d_{j-1})}e^{-2ik_{z,j}(d_{j-1}-d_{j-1})}] \tag{32}$$

$$AR_{j,o}^- = B_{j,o}^+ M_j \tilde{R}_{j|j+1} e^{2ik_{z,j}(d_j-d_{j-1})} \tag{33}$$

$$B_{j,o}^+ = \frac{A_{j,o}^+}{M_j} = \frac{iq}{\omega\varepsilon_0(2\pi)^3} \cdot a_{j-1|j}^{+,o} e^{i\frac{\omega}{v}d_{j-1}} \tag{34}$$

In above transformation, the relation of $M_j = 1 + M_j \tilde{R}_{j|j+1}\tilde{R}_{j|j-1}e^{2ik_{z,j}(d_j-d_{j-1})}$ is used. When calculating the field distribution in transmitted region $m$ ($m < j$) from the forward transition radiation at the $j-1|j$ interface, we only need to consider the second part relevant to $AR_{j,o}^-$ in equation (32). The calculation procedure is similar to that in *part two*. The field in region $m$ ($m < j$) can be expressed as follows:

$$E_{z,m}^R = C_{j|m}^- [e^{-ik_{z,m}(z-d_{j-1})} + \tilde{R}_{m|m-1}e^{+ik_{z,m}(z-d_{j-1})}e^{-2ik_{z,m}(d_{m-1}-d_{j-1})}] \tag{35}$$

where only $C_{j|m}^-$ is unknown. When $m = j-1$ in equation (35), at the $j|j-1$ interface with $z = d_{j-1}$, we have

$$C_{j|j-1}^- e^{-ik_{z,j-1}(d_{j-1}-d_{j-1})} = AR_{j,o}^- e^{-ik_{z,j}(d_{j-1}-d_{j-1})} \cdot S_{j|j-1} \tag{36}$$

$$S_{j|j-1} = \frac{T_{j|j-1}}{1 - R_{j-1|j}\tilde{R}_{j-1|j-2}e^{2ik_{z,j-1}(d_{j-1}-d_{j-2})}} \tag{37}$$

By following the operation in equation (36), one has

$$C_{j|j-2}^- e^{-ik_{z,j-2}(d_{j-2}-d_{j-1})} = C_{j|j-1}^- e^{-ik_{z,j-1}(d_{j-2}-d_{j-1})} \cdot S_{j-1|j-2} \tag{38}$$

$$C_{j|j-3}^- e^{-ik_{z,j-3}(d_{j-3}-d_{j-1})} = C_{j|j-2}^- e^{-ik_{z,j-2}(d_{j-3}-d_{j-1})} \cdot S_{j-2|j-3} \tag{39}$$

$$\ldots\ldots$$

$$C_{j|m+1}^- e^{-ik_{z,m+1}(d_{m+1}-d_{j-1})} = C_{j|m+2}^- e^{-ik_{z,m+2}(d_{m+1}-d_{j-1})} \cdot S_{m+2|m+1} \tag{40}$$

$$C_{j|m}^- e^{-ik_{z,m}(d_m-d_{j-1})} = C_{j|m+1}^- e^{-ik_{z,m+1}(d_m-d_{j-1})} \cdot S_{m+1|m} \tag{41}$$

From equations (36-41), one has

$$C_{j|m}^- e^{-ik_{z,m}(d_m-d_{j-1})} = \tilde{T}_{j|m}^{m<j} \cdot AR_{j,o}^- e^{-ik_{z,j}(d_j-d_{j-1})} \tag{42}$$

$$\tilde{T}_{j|m}^{m<j} = \prod_{n=m}^{n=j-1} S_{n+1|n} e^{-ik_{z,n+1}(d_n-d_{n+1})} \tag{43}$$

$\tilde{T}_{j|m}^{m<j}$ is the *generalized transmission coefficient* from region $j$ to region $m$ [49]. The field distribution in region $m$ ($m < j$) from the forward transition radiation at the $j-1|j$ interface can be solved by equation (42).

*Part four: Field distribution in region j from the backward transition radiation at the j|j + 1 interface*

For the backward transition radiation at the $j|j+1$ interface, from equation (8), we have the backward radiation factor as:

$$a_{j|j+1}^- = a_{j|j+1}^{-,o} e^{+ik_{z,j}d_j} e^{i\frac{\omega}{v}d_j} \tag{44}$$

From equation (7-1), the emitted field in region $j$ becomes:

$$E_{z,j}^R = A_{j,o}^- [e^{-ik_{z,j}(z-d_j)} + \tilde{R}_{j|j-1} e^{+ik_{z,j}(z-d_j)} \cdot e^{-2ik_{z,j}(d_{j-1}-d_j)}] \tag{45}$$

$$A_{j,o}^- = \frac{iq}{\omega \varepsilon_0 (2\pi)^3} \cdot a_{j|j+1}^{-,o} e^{i\frac{\omega}{v}d_j} \cdot M_j \tag{46}$$

Equation (45) can be equivalently transformed to:

$$E_{z,j}^R = B_{j,o}^- e^{-ik_{z,j}(z-d_j)} + BR_{j,o}^+ [e^{+ik_{z,j}(z-d_j)} + \tilde{R}_{j|j+1} e^{-ik_{z,j}(z-d_j)} e^{+2ik_{z,j}(d_j-d_j)}] \tag{47}$$

$$BR_{j,o}^+ = B_{j,o}^- M_j \tilde{R}_{j|j-1} e^{-2ik_{z,j}(d_{j-1}-d_j)} \tag{48}$$

$$B_{j,o}^- = \frac{A_{j,o}^-}{M_j} = \frac{iq}{\omega \varepsilon_0 (2\pi)^3} \cdot a_{j|j+1}^{-,o} e^{i\frac{\omega}{v}d_j} \tag{49}$$

*Part five: Field distribution in region m (m > j) from the backward transition radiation at the j|j + 1 interface*

By using the second part relevant to $BR_{j,o}^+$ in equation (47), the field distribution in region $m$ ($m > j$) from the backward transition radiation at the $j|j+1$ interface can be obtained. The field in region $m$ can be expressed as:

$$E_{z,m}^R = D_{j|m}^+ [e^{+ik_{z,m}(z-d_j)} + \tilde{R}_{m|m+1} e^{-ik_{z,m}(z-d_j)} \cdot e^{2ik_{z,m}(d_m-d_j)}] \tag{50}$$

The calculation procedure for the unknown factor $D_{j|m}^+$ is the same as that of $C_{j|m}^+$ in *part two*. After some algebra, we have

$$D_{j|m}^{+}e^{+ik_{z,m}(d_{m-1}-d_j)} = \tilde{T}_{j|m}^{m>j} \cdot BR_{j,o}^{+}e^{+ik_{z,j}(d_{j-1}-d_j)} \tag{51}$$

*Part six: Field distribution in region m (m < j) from the backward transition radiation at the j|j + 1 interface*

By using equation (45), the filed distribution in transmitted region $m$ $(m < j)$ from the backward transition radiation at the $j|j+1$ interface can be obtained. The field in region $m$ $(m < j)$ can be expressed as:

$$E_{z,m}^{R} = D_{j|m}^{-}[e^{-ik_{z,m}(z-d_j)} + \tilde{R}_{m|m-1}e^{+ik_{z,m}(z-d_j)}e^{-2ik_{z,m}(d_{m-1}-d_j)}] \tag{52}$$

The calculation procedure for the unknown factor $D_{j|m}^{-}$ is the same as that of $C_m^{-}$ in *part three*. After some algebra, we have

$$D_{j|m}^{-}e^{-ik_{z,m}(d_m-d_j)} = \tilde{T}_{j|m}^{m<j} \cdot A_{j,o}^{-}e^{-ik_{z,j}(d_j-d_j)} \tag{53}$$

This way, the field distribution in region $m$ $(m < j)$ from the backward transition radiation at the $j|j+1$ interface can be obtained by using equation (53).

*Part seven: Field distribution in region j from resonance transition radiation at multiple interfaces*

From above analysis, the total radiated field in region $j$ $(1 \leq j \leq N+1)$ can be a summation of two parts, i.e.

$$E_{z,j}^{R} = E_{z,j}^{R,+} + E_{z,j}^{R,-} \tag{54}$$

where $E_{z,j}^{R,+}$ and $E_{z,j}^{R,-}$ are attributed to the forward and backward transition radiations at each interface, respectively. Namely, $E_{z,j}^{R,+}$ is a summation of the emitted fields from the forward transition radiation at each interface transmitted into region $j$, and $E_{z,j}^{R,-}$ is a summation of the emitted fields from the backward transition radiation at each interface transmitted into region $j$. Therefore, we have

$$E_{z,j}^{R,+} = TR_j^{+} \to Region_j + \sum_{m=1}^{m=j-1} TR_m^{+} \to Region_j + \sum_{m=j+1}^{m=N+1} TR_m^{+} \to Region_j \tag{55}$$

$$TR_j^{+} \to Region_j = A_{j,o}^{+}[e^{+ik_{z,j}(z-d_{j-1})} + \tilde{R}_{j|j+1}e^{-ik_{z,j}(z-d_{j-1})} \cdot e^{2ik_{z,j}(d_j-d_{j-1})}] \tag{56}$$

$$\sum_{m=1}^{m=j-1} TR_m^{+} \to Region_j = \sum_{m=1}^{m=j-1} C_{m|j}^{+}[e^{+ik_{z,j}(z-d_{m-1})} + \tilde{R}_{j|j+1}e^{-ik_{z,j}(z-d_{m-1})} \cdot e^{2ik_{z,j}(d_j-d_{m-1})}] \tag{57}$$

$$\sum_{m=j+1}^{m=N+1} TR_m^+ \to Region_j = \sum_{m=j+1}^{m=N+1} C_{m|j}^- [e^{-ik_{z,j}(z-d_{m-1})} + \tilde{R}_{j|j-1} e^{+ik_{z,j}(z-d_{m-1})} e^{-2ik_{z,j}(d_{j-1}-d_{m-1})}] \quad (58)$$

$$E_{z,j}^{R,-} = TR_j^- \to Region_j + \sum_{m=1}^{m=j-1} TR_m^- \to Region_j + \sum_{m=j+1}^{m=N+1} TR_m^- \to Region_j \quad (59)$$

$$TR_j^- \to Region_j = A_{j,o}^- [e^{-ik_{z,j}(z-d_j)} + \tilde{R}_{j|j-1} e^{+ik_{z,j}(z-d_j)} \cdot e^{-2ik_{z,j}(d_{j-1}-d_j)}] \quad (60)$$

$$\sum_{m=1}^{m=j-1} TR_m^- \to Region_j = \sum_{m=1}^{m=j-1} D_{m|j}^+ [e^{+ik_{z,j}(z-d_m)} + \tilde{R}_{j|j+1} e^{-ik_{z,j}(z-d_m)} \cdot e^{2ik_{z,j}(d_j-d_m)}] \quad (61)$$

$$\sum_{m=j+1}^{m=N+1} TR_m^- \to Region_j = \sum_{m=j+1}^{m=N+1} D_{m|j}^- [e^{-ik_{z,j}(z-d_m)} + \tilde{R}_{j|j-1} e^{+ik_{z,j}(z-d_m)} e^{-2ik_{z,j}(d_{j-1}-d_m)}] \quad (62)$$

This way, we can obtain the field distribution of resonance transition radiation from multiple interfaces in each region by using equations (55-62). For regions $j = 1$ and $j = N + 1$, the radiated fields only propagate along the $-z$ and $+z$ directions, respectively. When $j = 1$, we have

$$E_{z,1}^{R,-} = \frac{iq}{\omega \varepsilon_0 (2\pi)^3} \cdot a_1 \cdot e^{-ik_{z,1}z} \quad (63)$$

$$a_1 = \frac{\omega \varepsilon_0 (2\pi)^3}{iq} \left( A_{1,o}^- e^{+ik_{z,1}d_j} + \sum_{m=2}^{m=N+1} D_{m|1}^- e^{+ik_{z,1}d_m} + \sum_{m=2}^{m=N+1} C_{m|1}^- e^{+ik_{z,1}d_{m-1}} \right) \quad (64)$$

When $j = N + 1$, we have

$$E_{z,N+1}^{R,+} = \frac{iq}{\omega \varepsilon_0 (2\pi)^3} \cdot a_{N+1} \cdot e^{+ik_{z,N+1}z} \quad (65)$$

$$a_{N+1} = \frac{\omega \varepsilon_0 (2\pi)^3}{iq} \left( A_{N+1,o}^+ e^{-ik_{z,N+1}d_N} + \sum_{m=1}^{m=N} C_{m|N+1}^+ e^{-ik_{z,N+1}d_{m-1}} + \sum_{m=1}^{m=N} D_{m|N+1}^+ e^{-ik_{z,N+1}d_m} \right) \quad (66)$$

It is worthy to note that the form of equations (63,65) is the same as that of equations (7-1,7-2).

### Supplementary note 3: Angular spectral energy density of resonance transition radiation

The angular spectral energy density of resonance transition radiation in the forward and backward directions can be analytically obtained, where the calculation procedure is briefly introduced below; see more details in our previous work [39]. One can obtain the total energy $W_1$ radiated by a charged particle into region 1 (top air region), i.e. backwards relative to its motion, by integrating the emitted field energy density over all space. For a long time $t$, the radiated wave-train is already at a great distance to the interface and then separated from the charge's intrinsic field. If the origin is moved along the axis into the region of the radiated wave-train, the integration with respect to $z$ can be taken from $-\infty$ to $+\infty$, since the field is attenuated in both directions [39]. For freely propagating waves, since the electric and magnetic energy densities are equal, one has

$$W_1 = \int dxdy \int_{-\infty}^{+\infty} dz \cdot \varepsilon_1 |\bar{E}_1^R(\bar{r},t)|^2 \tag{67}$$

$$|\bar{E}_1^R(\bar{r},t)|^2 = \int \bar{E}_{1|\bar{\kappa}_\perp,\omega}^R(z) \cdot \bar{E}_{1|\bar{\kappa}_\perp',\omega'}^{R\ *}(z) e^{i[(\bar{\kappa}_\perp - \bar{\kappa}_\perp')\cdot \bar{r}_\perp - (\omega-\omega')t]} d\bar{\kappa}_\perp d\bar{\kappa}_\perp' d\omega d\omega' \tag{68}$$

Substituting equation (68) into equation (67) and integrating over $dxdyd\bar{\kappa}_\perp'\,dzd\omega'$ in equation (67) gives

$$W_1 = 2\int_0^{+\infty}\int \varepsilon_1 |a_1|^2 \left(\frac{q}{\omega \varepsilon_0 (2\pi)^3}\right)^2 \frac{\omega^2}{c\kappa_\perp^2} \sqrt{\varepsilon_{1r}} \sqrt{1 - \frac{\kappa_\perp^2 c^2}{\omega^2 \varepsilon_{1r}}} (2\pi)^3 d\bar{\kappa}_\perp d\omega \tag{69}$$

For emitted photons, the integration over $d\bar{\kappa}_\perp$ is to be taken over the range $\kappa_\perp^2 < \frac{\omega^2}{c^2}\varepsilon_{1r}$. We use the angle $\theta$ between the radiation wave vector $\bar{k}_1 = (\bar{\kappa}_\perp, \hat{z}k_{1z})$ and the direction of the vector $-\bar{v}$, so that $\theta = 0$ represents radiation in the opposite direction of the particle's motion. Then we can express $\kappa_\perp = \frac{\omega}{c}\sqrt{\varepsilon_{1r}}\sin\theta$. A further change from integration over $d\bar{\kappa}_\perp$ to one over $2\pi\kappa_\perp d\kappa_\perp = 2\pi(\frac{\omega^2}{c^2}\varepsilon_{1r})\sin\theta\cos\theta d\theta$ in equation (69) gives

$$W_1 = \int_0^{+\infty}\int_0^{\pi/2} U_1(\omega,\theta)\cdot(2\pi\sin\theta)d\theta\,d\omega \tag{70}$$

$$U_1(\omega,\theta) = \frac{\varepsilon_{1r}^{3/2} q^2 \cos^2\theta}{4\pi^3 \varepsilon_0 c \sin^2\theta}|a_1|^2 \tag{71}$$

Equation (71) is the backward angular spectral energy density, which shows the distribution of the radiation as a function of frequency and angle. With equation (64), the backward angular spectral energy density in equation (71) can be obtained straightforwardly. By expressing the total energy as $W_1 = \int_0^\infty W_1(\omega)d\omega$, one has the energy spectrum of resonance transition radiation in the backward direction as

$$W_1(\omega) = \int_0^{\pi/2} U_1(\omega,\theta)\cdot(2\pi\sin\theta)d\theta \tag{72}$$

By following the calculation procedure above, the forward angular spectral energy density and the energy spectrum of resonance transition radiation in region $N + 1$ (bottom air region) can also be obtained as:

$$U_{N+1}(\omega, \theta) = \frac{\varepsilon_{r,N+1}^{3/2} q^2 \cos^2\theta}{4\pi^3 \varepsilon_0 c \sin^2\theta} |a_{N+1}|^2 \tag{73}$$

$$W_{N+1}(\omega) = \int_0^{\pi/2} U_{N+1}(\omega, \theta) \cdot (2\pi \sin\theta) d\theta \tag{74}$$

With equation (66), the forward angular spectral energy density in equation (73) can be calculated straightforwardly. Note that the definitions of the radiation angle for the backward radiation in equations (71) and the forward radiation in equation (73) are different. For the backward radiation in equations (71), the radiation angle as stated above is the angle between the radiation wave vector $\bar{k}_1 = (\bar{\kappa}_\perp, \hat{z} k_{1z})$ and the vector $-\bar{v}$; namely $\theta = \theta_B$ in Fig. 1. For the forward radiation in equations (73), the radiation angle is the angle between the radiation wave vector $\bar{k}_1 = (\bar{\kappa}_\perp, \hat{z} k_{1z})$ and the vector $+\bar{v}$; namely $\theta = \theta_F$ in Fig. 1.

The total emitted photon spectrum is $W(\omega) = W_1(\omega) + W_2(\omega)$. The total energy of emitted photons into the top and bottom air regions can be obtained by integrating over frequency.

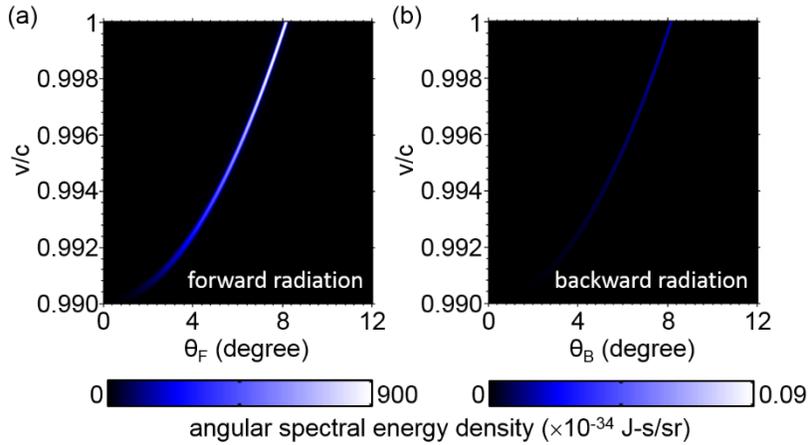

**Supplementary Figure 3 | Cherenkov radiation from an isotropic slab.** The isotropic slab with a refractive index of $n = 1.01$ has a thickness of 2 mm. The structure is air-dielectric slab-air. (a-b) Angular spectral energy density of forward (backward) radiation in the bottom (top) air region at the working frequency (corresponding wavelength in air is $\lambda = 700$ nm).

*Angular spectral energy density of transition radiation from an isotropic slab*

Supplementary Figure 3 shows the angular spectral energy density of transition radiation from an isotropic slab with a hypothetical refractive index of $n = 1.01$. The highly directional radiation in Supplementary Fig. 3 (a) shows the relation between the Cherenkov angle and the particle velocity (denoted as the Cherenkov relation below). From the Cherenkov relation, the isotropic slab will have the forward Cherenkov radiation when the particle velocity is in the range of $[0.99c \ c]$. The corresponding range for the Cherenkov angle in the air region is [0 8.1] degree. The angular spectral energy density in the forward direction in Supplementary Fig. 3(a) is ~$10^5$ times of that in the backward direction in Supplementary Fig. 3(b).

*Forward Cherenkov relation of photonic crystals for particle identification with other velocity or momentum*

Here we demonstrate the forward Cherenkov relation from an alternative photonic crystal that can enable the identification of particles with the extremely high velocity or momentum, as shown in Supplementary Fig. 4. The highly directional radiation in Supplementary Figs. 4(a,b) shows the Cherenkov relation of the photonic crystal. The corresponding ranges of the particle velocity and the Cherenkov angle in the air region are $[0.9995c, c]$ and [0 2.11] degree, respectively. The angular spectral energy density in the forward direction in Supplementary Fig. 4(a) is ~100 times larger than that in the backward direction in Supplementary Fig. 4(b). Therefore, the resonance transition radiation from the photonic crystal here is equivalently to the forward Cherenkov radiation. By applying the Cherenkov relation in Supplementary Fig. 4(a), Supplementary Fig. 4(c) shows the relation between the particle momentum and the Cherenkov angle for four kinds of particles with different masses. For different particles with the same momentum being as high as 29.7 GeV/c, the corresponding Cherenkov angle is 2.12°, 2.09°, 1.79°, and 0° for the electron, pion, kaon, and proton, respectively. These different Cherenkov angles can make different particles to be effectively distinguished.

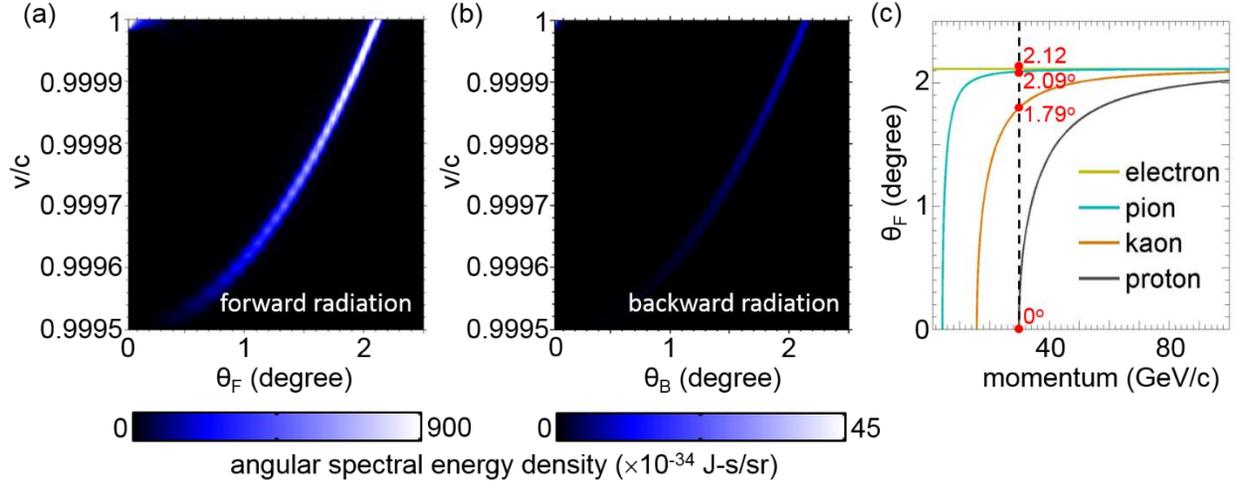

**Supplementary Figure 4 | Controlling the forward Cherenkov angle by photonic crystals.** For the studied photonic crystal, the thickness of the unit cell is $d_{unit} = 1.0146\lambda$; within the unit cell, the thicknesses for the two dielectric constituents are $d_1 = 0.3 d_{unit}$ and $d_2 = 0.7 d_{unit}$, respectively. The total thickness of the photonic crystal is 50 mm. (a-b) Angular spectral energy density of forward (backward) radiation in the bottom (top) air region. (c) Cherenkov angles versus the particle momentum for four kinds of particles. These results adopt the relation between the particle velocity and the Cherenkov angle in (a), and translate the velocity to the momentum using the masses of different charged particles.

## Supplementary note 4: Resonance transition radiation from a uniaxial system

The calculation procedure of resonance transition radiation in a uniaxial system is similar to that in supplementary notes 1-3, which is briefly summarized below. We only consider resonance transition radiation in an isotropic system in supplementary notes 1-3, where all regions are composed of isotropic materials. Actually, resonance transition radiation from multiple interfaces can also be analytically solved when the material in each becomes uniaxial with the relative permittivity being $\bar{\bar{\varepsilon}}_{r,j} = [\varepsilon_{r,\perp,j}, \varepsilon_{r,\perp,j}, \varepsilon_{r,z,j}]$ for region $j$ [35,36]. When $\varepsilon_{r,\perp,j} = \varepsilon_{r,z,j} = \varepsilon_{r,j}$, the material in region $j$ becomes isotropic again.

For transition radiation from a single interface, when the two regions becomes uniaxial, the freely radiated fields can be still be expressed by equations (7-9), but the two radiation facts of $a_{j|j+1}^{-,o}$ in equation (10) and $a_{j|j+1}^{+,o}$ in equation (10) are changed to:

$$a_{j|j+1}^{-,o} = \frac{\kappa_\perp^2 c^2}{\omega^2} \cdot \frac{-v}{c} \cdot \frac{\varepsilon_{r,\perp,j}\varepsilon_{r,\perp,j+1}}{\varepsilon_{r,z,j}} \cdot \frac{\frac{1-\frac{v}{c}\frac{k_{z,j+1}}{\omega/c}}{\varepsilon_{r,z,j+1}(1-\frac{v^2}{c^2}\varepsilon_{r,\perp,j+1}+\frac{\kappa_\perp^2 v^2}{\omega^2}\frac{\varepsilon_{r,\perp,j+1}}{\varepsilon_{r,z,j+1}})} - \frac{1-\frac{v}{c}\frac{k_{z,j+1}}{\omega/c}\frac{\varepsilon_{r,\perp,j}}{\varepsilon_{r,\perp,j+1}}}{\varepsilon_{r,z,j}(1-\frac{v^2}{c^2}\varepsilon_{r,\perp,j}+\frac{\kappa_\perp^2 v^2}{\omega^2}\frac{\varepsilon_{r,\perp,j}}{\varepsilon_{r,z,j}})}}{\varepsilon_{r,\perp,j}\frac{k_{z,j+1}}{\omega/c}+\varepsilon_{r,\perp,j+1}\frac{k_{z,j}}{\omega/c}} \quad (75)$$

$$a_{j|j+1}^{+,o} = \frac{\kappa_\perp^2 c^2}{\omega^2} \cdot \frac{+v}{c} \cdot \frac{\varepsilon_{r,\perp,j}\varepsilon_{r,\perp,j+1}}{\varepsilon_{r,z,j+1}} \cdot \frac{\frac{1+\frac{v}{c}\frac{k_{z,j}}{\omega/c}}{\varepsilon_{r,z,j}(1-\frac{v^2}{c^2}\varepsilon_{r,\perp,j}+\frac{\kappa_\perp^2 v^2}{\omega^2}\frac{\varepsilon_{r,\perp,j}}{\varepsilon_{r,z,j}})} - \frac{1+\frac{v}{c}\frac{k_{z,j}}{\omega/c}\frac{\varepsilon_{r,\perp,j+1}}{\varepsilon_{r,\perp,j}}}{\varepsilon_{r,z,j+1}(1-\frac{v^2}{c^2}\varepsilon_{r,\perp,j+1}+\frac{\kappa_\perp^2 v^2}{\omega^2}\frac{\varepsilon_{r,\perp,j+1}}{\varepsilon_{r,z,j+1}})}}{\varepsilon_{r,\perp,j}\frac{k_{z,j+1}}{\omega/c}+\varepsilon_{r,\perp,j+1}\frac{k_{z,j}}{\omega/c}} \quad (76)$$

In addition, it is worthy to note that the component of the wavevectors for the radiated fields along the $z$ direction are changed to $k_{z,j} = \sqrt{\frac{\varepsilon_{r,\perp,j}\omega^2}{c^2} - \kappa_\perp^2 \frac{\varepsilon_{r,\perp,j}}{\varepsilon_{r,z,j}}}$ and $k_{z,j+1} = \sqrt{\frac{\varepsilon_{r,\perp,j+1}\omega^2}{c^2} - \kappa_\perp^2 \frac{\varepsilon_{r,\perp,j+1}}{\varepsilon_{r,z,j+1}}}$, respectively.

For resonance transition radiation from multiple interfaces in a uniaxial system, the field distribution in each region can still be expressed by equations (54-66). However, the reflection and transmission coefficients at each interface (see the text below equation (22-2)) are changed to $R_{j|j+1} = -R_{j+1|j} = \frac{\frac{k_{z,j}}{\varepsilon_{r,\perp,j}}-\frac{k_{z,j+1}}{\varepsilon_{r,\perp,j+1}}}{\frac{k_{z,j}}{\varepsilon_{r,\perp,j}}+\frac{k_{z,j+1}}{\varepsilon_{r,\perp,j+1}}}$, $T_{j|j+1} = \frac{2\frac{k_{z,j}}{\varepsilon_{r,\perp,j}}}{\frac{k_{z,j}}{\varepsilon_{r,\perp,j}}+\frac{k_{z,j+1}}{\varepsilon_{r,\perp,j+1}}} \cdot \frac{\varepsilon_{r,z,j}}{\varepsilon_{r,z,j+1}}$ and $T_{j+1|j} = \frac{2\frac{k_{z,j+1}}{\varepsilon_{r,\perp,j+1}}}{\frac{k_{z,j+1}}{\varepsilon_{r,\perp,j+1}}+\frac{k_{z,j}}{\varepsilon_{r,\perp,j}}} \cdot \frac{\varepsilon_{r,z,j+1}}{\varepsilon_{r,z,j}}$, respectively.

Since we consider the top and bottom regions both to be the isotropic air region in this work, the backward and forward angular spectral energy densities of resonance transition radiation in a uniaxial system can still be expressed by equations (71,73), respectively.

*Loss influence on the angular spectral energy density of transition radiation from a uniaxial slab*

Recently, under the ideal lossless assumption, the uniaxial metal-based metamaterials are proposed to control the Cherenkov relation [14]. Since losses are unavoidably in the metal-based systems, Supplementary Fig. 5 investigate the loss influence on the angular spectral energy density of the forward Cherenkov radiation from a uniaxial slab; see the structure in Supplementary Fig. 5(a). For brevity, we consider the permittivity of the uniaxial slab to be equal to the permittivity of the uniaxial metal-based metamaterials in Ref.[14]. When under the ideal lossless assumption, the Cherenkov relation can be clearly seen from the forward angular spectral energy density in Supplementary Fig. 5(b). However, when the realistic loss is considered, the Cherenkov relation is severely destroyed, as shown in Supplementary Fig. 5(c). In addition, the value of the forward angular spectral energy density for the lossless case in Supplementary Fig. 5(b) is over 200 times larger than that for the lossy case in Supplementary Fig. 5(c). Therefore, the realistic loss is detrimental to the performance of Cherenkov detectors, and it shall be judicious to control the Cherenkov relation *only* in lossless systems.

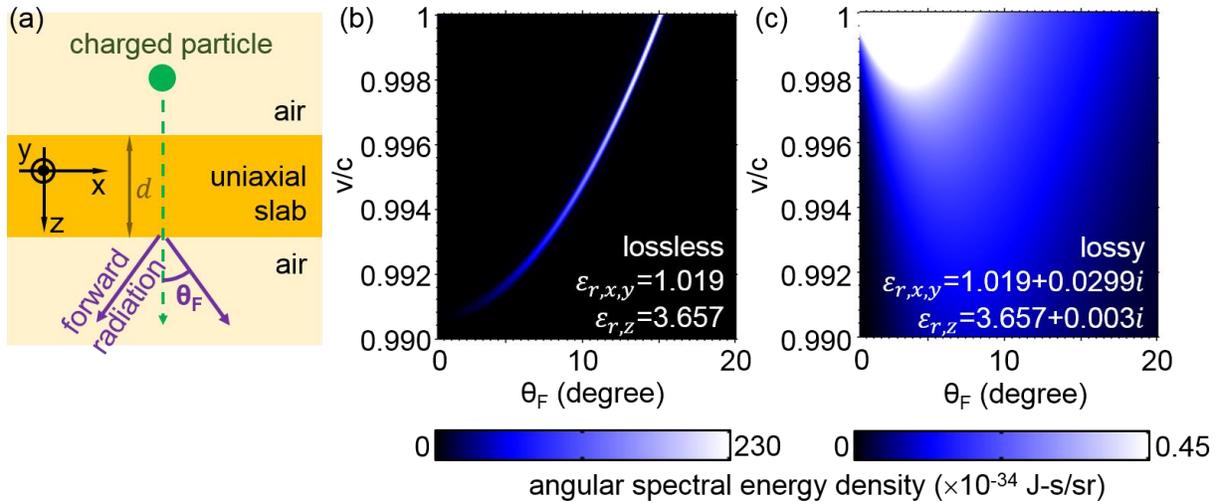

**Supplementary Figure 5 | Loss influence on the forward Cherenkov radiation from a uniaxial slab.** (a) Structural schematic. The uniaxial slab with a thickness of $d = 2$ mm has a relative permittivity of $[\varepsilon_{r,x}, \varepsilon_{r,y}, \varepsilon_{r,z}]$, where $\varepsilon_{r,x} = \varepsilon_{r,y}$. (b-c) Angular spectral energy density of the forward radiation in the top air region from a (b) lossless and (c) lossy uniaxial slab. For the lossy case in (c), the relative uniaxial permittivity, which is adopted from Ref.[14], stands for a metal-based uniaxial metamaterials. For the ideal lossless case in (b), the imaginary part of the uniaxial permittivity is neglected.

## Supplementary note 5: Band structure and isofrequency contour of one-dimensional photonic crystals

The dispersion of 1D photonic crystal has been extensively studied and can be calculated by matching the boundary condition with the help of Bloch's theorem (or Bloch-periodic boundary condition) [31]. In this work, the 1D photonic crystal is composed by two different isotropic materials along the $z$ direction, where their relative permittivities are $\varepsilon_{r,1}$ and $\varepsilon_{r,2}$, respectively. For a periodic unit cell with a total thickness of $d = d_1 + d_2$, the thickness of these two materials are $d_1$ and $d_2$, respectively. The dispersion of the 1D photonic crystal for TM waves can be expressed as:

$$\cos(k_z d) = \cos(k_{z1} d_1)\cos(k_{z2} d_2) - \frac{1}{2}\left(\frac{Y_1}{Y_2} + \frac{Y_2}{Y_1}\right)\sin(k_{z1} d_1)\sin(k_{z2} d_2) \tag{77}$$

$$Y_1 = \omega \varepsilon_0 \varepsilon_{r,1}/k_{z1}; \quad Y_2 = \omega \varepsilon_0 \varepsilon_{r,2}/k_{z2} \tag{78}$$

In above equations, $k_{z,1} = \sqrt{\frac{\varepsilon_{r,1}\omega^2}{c^2} - \kappa_\perp^2}$ and $k_{z,2} = \sqrt{\frac{\varepsilon_{r,2}\omega^2}{c^2} - \kappa_\perp^2}$; $\kappa_\perp$, same as that in supplementray notes 1-4, is the magnitude of the component of the wavevector parallel to the interface. When $\kappa_\perp = 0$, equation (77) characterizes the relation between the angular frequency ω and the component of the wavevector along the $z$ direction (i.e. $k_z$). By solving this case, the band structure for the 1D photonic crystal can be obtained. When the angular frequency is a constant with $\omega = \omega_0$, equation (77) characterizes the relation between $k_z$ and $\kappa_\perp$. By solving this case, the isofrequency contour of the 1D photonic crystal at $\omega = \omega_0$ can be obtained.

## Supplementary References